\documentclass{JHEP}
\usepackage{amsmath,amssymb}
\newcommand{\ii}{\mathrm{i}}
\newcommand{\pd}{\partial}
\newcommand{\e}{\mathrm{e}}
\newcommand{\const}{\mathrm{const}}
\newcommand{\half}{\frac{1}{2}}
\newcommand{\I}{\mathbb{I}}
\newcommand{\A}{\mathcal{A}}
\newcommand{\B}{\mathcal{B}}

\newcommand{\F}{{\mathcal F}}

\newcommand{\hh}{\mathcal{H}}
\title{Noncommutative Tachyonic Solitons. Interaction with Gauge
Field\thanks{Work supported by RFBR grant \# 99-01-00190, INTAS
grant \# 950681, and Scientific School support grant
96-15-0628}}%
\author{Corneliu Sochichiu\\The Abdus Salam ICTP, strada Costiera 11, 34100 Trieste,
Italy\\ Institutul de Fizic\u a Aplicat\u a
A\c S, str. Academiei, nr. 5, Chi\c sin\u au, MD2028 MOLDOVA\\
and\\ Bogoliubov Laboratory of Theoretical Physics, Joint
Institute for Nuclear Research, 141980 Dubna, Moscow Reg., RUSSIA
\\
E-mail: \email{sochichi@thsun1.jinr.ru}}%
\preprint{\hepth{0007217}}
\abstract{We show that in the presence of U(1) non-commutative YM
interaction the noncommutative tachyonic system exhibits solitonic
solutions (even) for finite value of non commutativity parameter
$\theta$.}
\begin{document}

\section{Introduction}
% ----------------------------------------------------------------
The recent interest to the non commutative models have its source
in their deep relation to the string theory,
\cite{Witten:1986cc,Seiberg:1999vs,Witten:2000nz}.

For example, the decaying modes of non-BPS unstable branes in IIA
and IIB string theories can be described in terms of
noncommutative tachyonic solitons
\cite{Gopakumar:2000zd,Dasgupta:2000ft,Harvey:2000jt}. This
description is based on solutions found in Ref.
\cite{Gopakumar:2000zd}, when the noncommutativity parameter
$\theta$ goes to infinity. In this limit one can neglect the
kinetic term in the equations of motion, and this allows one
obtaining nontrivial solutions depending only on the minima of the
tachyonic potential. These solutions are localised in a finite
region of the space and can be interpreted as solitons
\cite{Gopakumar:2000zd}. So far, the tachyonic system was
considered on the noncommutative space with constant (and large)
noncommutativity parameter $\theta$. This $\theta$ corresponds to
constant B-field on the brane. In order to get the desired value
of $\theta$, one has to take the limit, when B-field is much
bigger than metric but small enough to give the large value of
$\theta$, \cite{Harvey:2000jt}.

From the other hand one can go beyond the approximation of
constant $\theta$  (or $B$-field) and allow also dynamics for this
parameter. We put forward the idea to describe the
\emph{dynamical} noncommutativity through the \emph{U(1)
noncommutative Yang--Mills model} arising from IKKT matrix model
\cite{Ishibashi:1996xs}, at $N=\infty$,
\cite{Connes:1998cr,Sochichiu:2000ud,Sochichiu:2000fs,%
Sochichiu:2000bg}.

In this model the noncommutativity parameter $\theta=B^{-1}$
arises as the r.h.s of the commutator of the solution
$\A^{(0)}_{\mu}=p_{\mu}$ to the equations of motion:
\begin{equation}\label{heisen}
[p_{\mu},p_{\nu}]=\ii B_{\mu\nu}.
\end{equation}

This solution can be seen as one generating a flat noncommutative
space-time in Connes' approach \cite{Connes:2000by}. The generic
configuration of noncommutative gauge fields $\A_{\mu}$ can be
described as noncommutative functions on $x^{\nu}=\theta^{\mu\nu}
p_{\nu}$, which are subject to the Moyal product (or star
product) defined as follows,
\begin{equation}\label{star}
  \A *\B(x)=
  \e^{\frac{\ii}{2}\theta^{\mu\nu}\pd_\mu\pd_\nu'}\A(x) \B(x')
  \big|_{x'=x},
\end{equation}
where the $\pd_\mu$ and $\pd_\mu'$ denote derivatives with respect
to $x$ and $x'$ respectively. In Ref. \cite{Sochichiu:2000bg} we
have shown that in the case of noncommutative gauge field the
algebra (\ref{star}) is equivalent to its two dimensional
reduction.

Extending the above case one can identify the dynamical
noncommutativity parameter with the strength tensor of the gauge
field $\A_{\mu}$.

In what follows, we are going to show that allowing the
noncommutative parameter to be dynamical and not just constant one
makes possible finding the solitonic solutions without taking any
limit.

The plan of the paper is as follows. First, we introduce the model
describing the tachyonic system interacting with the gauge field
$\A_{\mu}$. The interaction is introduced as the gauging
noncommutative U(1) gauge symmetry from ``nothing''. After that we
consider a class of solitonic ans\"{a}tze for equations of motion
and find the consistent one. Finally, we discuss the results.

% ----------------------------------------------------------------
\section{The model}
% ----------------------------------------------------------------
In Ref. \cite{Dasgupta:2000ft}, it was claimed that in the
presence of B-field an unstable Dp-brane is described by the
following noncommutative tachyonic action,
\begin{equation}\label{tachyon}
    S=\int d^{p+1}x \left(\frac12 \pd_\mu\phi\pd_\mu\phi-V(*\phi)
    \right),
\end{equation}
where the star reminds that all the products are taken to be star
ones, and $V(* \phi)$ is the tachyonic potential
\cite{Sen:1999md,Bergshoeff:2000dq,Garousi:2000tr}. Generally,
the particular form of $V(*\phi)$ is not known, except the fact
that it has negative slope at the origin (this is why it is
tachyonic) and nontrivial minima outside origin, where it is
negative. Since the negative energy of the tachyonic potential
can compensate the positive brane tension this allows the decaying
of branes to nothing \cite{Dasgupta:2000ft,Harvey:2000jt}.

The action (\ref{tachyon}) can be considered as one possessing
``invariance'' under global gauge symmetry given by constant
noncommutative unitary transformations,
\begin{equation}\label{glob_sym}
   \phi\to U^\dag*\phi*U, \qquad U^\dag*U=1
\end{equation}
since $U=\e^{\ii \varphi}=\const$, the star products in the above
equation are equivalent to ordinary products.

The fields are represented as operators on a Hilbert space $\hh$,
and symmetry (\ref{glob_sym}), just represents the phase
invariance of quantum states. Although, the action of this
symmetry on fields $\phi$ is trivial, ($U^\dag \phi U\equiv
\phi$), due to noncommutativity one gets non trivial gauging of
this action. Indeed, non constant $U$ is the picture changing
unitary operator which, generally, do not leave $\phi$ invariant.
The action (\ref{tachyon}), is not invariant with respect to this
transformation of the field $\phi$. The invariance of the action,
however, can be restored by the gauging derivatives in
(\ref{tachyon}). As in conventional theory, the gauging is
obtained by substitution of ordinary derivatives by covariant
ones, the only difference being that the gauge fields are also
noncommutative,
\begin{equation}\label{nabla}
   \pd_\mu\phi\to\nabla_\mu\phi\equiv\pd_\mu\phi+\ii [\A_\mu,\phi],
\end{equation}
we remind that the commutator in (\ref{nabla}) is computed using
the star product (\ref{star}).

Introducing also the Yang--Mills part for the field $\A$, one has
for the total action,
\begin{equation}\label{action}
   S=\int d^{p+1}x \left(\frac12 \nabla_\mu\phi\nabla_\mu\phi-V(*\phi)
   -\frac{1}{4g^2}\F_{\mu\nu}^2\right),
\end{equation}
where $\F_{\mu\nu}$ is the gauge field strength,
\begin{equation}\label{f}
   \F_{\mu\nu}=\ii \left(\pd_\mu\A_\nu-\pd_\nu\A_\mu+\ii
   [\A_\mu,\A_\nu]\right),
\end{equation}
here, as well as in the consequent equations all the products are
the star ones, note also that the gauge field $\A_\mu$ is in the
Hermitian form while the gauge strength $\F_{\mu\nu}$ is an
anti-Hermitian one.

Due to the noncommutativity one can eliminate the derivatives from
the kinetic terms of action (\ref{action}) by shifting the gauge
field $\A_\mu$ as follows,
\begin{equation}\label{shift}
   \A_\mu\to p_\mu+\A_\mu,
\end{equation}
where $p_\mu=\theta^{-1}_{\mu\nu}x^\nu\equiv B_{\mu\nu}x^\nu$.

Indeed, the action (\ref{action}) depends on $\A$ only through the
covariant derivatives, while the covariant derivative of an
arbitrary function $f$ can be represented as,
\begin{equation}
   \nabla_\mu f=\pd_\mu+\ii[\A,f]=\ii [(p_\mu+\A_\mu),f],
\end{equation}
from which it immediately follows that all covariant derivatives
in action (\ref{action}) are replaced by commutators $[\A_\mu,f]$.

As a result the action (\ref{action}) looks as follows,
\begin{equation}\label{action_a}
  S=\int d^{p+1}x \left(\frac12 [\A_\mu,\phi]^2-V(\phi)
  -\frac{1}{4}[\A_\mu,\A_\nu]^2\right),
\end{equation}
where we also put $g=1$.

Action (\ref{action_a}) produce the following equations of motion,
\begin{align}\label{em:phi}
   &\frac{\pd V(\phi)}{\pd\phi}-[\A_\mu,[\A_\mu,\phi]]=0,\\
   \label{em:a}
   &[\A_\mu,[\A_\mu,\A_\nu]]+[\phi,[\A_\mu,\phi]]=0.
\end{align}

An interpretation which can be given to the action
(\ref{action_a}), is one of a tachyonic field $\phi$ living on a
noncommutative space-time generated by operators $\A_\mu$, when
one interpret them in the sense of Connes approach as the
(noncommutative) space-time position operators
\cite{Sochichiu:2000ud}. Indeed, in the case when gauge fields
$\A_\mu$ form an irreducible set (i.e. the only function
commuting with all $\A_\mu$ is the constant one), one can express
the fields in terms of operator functions $\phi=\phi(\A)$.

% ----------------------------------------------------------------
\section{Solitonic solutions}
% ----------------------------------------------------------------
Consider the equations of motion (\ref{em:phi},\ref{em:a}), and
let us look for the static solutions, i.e. ones commuting with
$\A_0$,
\begin{equation}\label{static}
   [\A_0,\A_i]=[\A_0,\phi]=0, \qquad i=1,\dots, p.
\end{equation}

This truncates the equation of motion to the following form,
\begin{align}\label{static:phi}
   &\frac{\pd V(\phi)}{\pd\phi}-[\A_i,[\A_i,\phi]]=0,\\
   \label{static:a}
   &[\A_i,[\A_i,\A_j]]+[\phi,[\A_j,\phi]]=0,
\end{align}
where Latin indices run through space-like directions of the
brane $i=1,\dots,p$.

Let us find solutions to eqs. (\ref{static:phi},\ref{static:a}).
In the case if there were no commutator term in equation
(\ref{static:phi}), the solution to this equation would be given
by a finite sum \cite{Gopakumar:2000zd},
\begin{equation}\label{sol}
   \phi=\sum_I a_I\Phi_I,
\end{equation}
where $a_I$ are the minima of the tachyonic potential $V(a)$,
treated as an ordinary (commutative) function, and $\Phi_I$ are
mutually orthogonal projectors to finite dimensional subspaces of
the Hilbert space $\hh$,
\begin{equation}
   \Phi_I\Phi_J=\delta_{IJ}\Phi_J,
\end{equation}
where no sum is assumed over repeated $J$ in the above equation.

Let us return back to the truncated equations of motion
(\ref{static:phi},\ref{static:a}). The above arguments apply also
to our case if the term with double commutator of $\A_i$ with
$\phi$ vanishes, i.e.,
\begin{equation}\label{laplace}
   [\A_i,[\A_i,\phi]]=0.
\end{equation}
The equation (\ref{laplace}), is the Laplace equation in the
presence of the gauge field $\A_\mu$.

In what follows our strategy will consist in solving the simple
tachyonic equation
\begin{equation}\label{tachyonic}
    \frac{\pd V(\phi)}{\pd \phi}=0,
\end{equation}
and after that finding the gauge field backgrounds which satisfy
the remaining equations of motion common with the condition
(\ref{laplace}). We are not going to find the general solution to
(\ref{laplace}), but instead propose a number of possible
ans\"{a}tze, without claiming to enumerate all the possibilities.
The ans\"{a}tze are as follows,
\begin{align*}
   \text{i)}&\qquad [\A_i,\phi_0]=0, \\
   \text{ii)}&\qquad [\A_i,\phi_0]=c_i,\\
   \text{iii)}&\qquad [\A_i,\phi_0]=c\A_i,\\
   \text{iv)}&\qquad [\A_i,\phi_0]=\pi_i \phi_0,
\end{align*}
where $\phi_0$ is the solitonic solution to eq.
(\ref{static:phi}), and $c_i$, $c$ and $\pi_i$ are arbitrary
constants.

It is worthwhile to note that in the last case iv) for $\pi_i\neq
0$ the  eq. (\ref{tachyonic}) is not satisfied but reduces to the
form,
\begin{equation}\label{tilda:eq}
   \frac{\pd \widetilde{V}}{\pd \phi}=0,
\end{equation}
where the modified tachyonic potential $\tilde {V}$ is given by,
\begin{equation}
   \widetilde{V}=V+\half\pi^2\phi.
\end{equation}

In this case $\phi_0$ is a solution to (\ref{tilda:eq}), given by
(\ref{sol}), where $a_I$ should be substituted by $\tilde{a}_I$
which now are minima of $\tilde{V}$.

As it is not difficult to show the ans\"{a}tze iii)-iv) for
nonzero $c$, $c_i$ and $\pi_i$ lead for $\A_i$ to trivial
solutions only. E.g. in the case ii) multiplying from left and
right by a factor $(1-\phi_0)$ gives the following identity (for
simplicity we assume that potential $V(a)$ has the only nontrivial
minimum),
\begin{equation}
   0=c_i(1-\phi_0),
\end{equation}
which is satisfied for either trivial $\phi_0$ or trivial $c_i$.
In an analogous way one can show that iii) and iv) for nonzero
$c$ and $\pi_i$ are consistent only for trivial $\A_i$,
respectively, trivial $\phi_0$.

The triviality of $\A_i$ can be interpreted as the space being
collapsed to a point.

In what follows consider in more details the remaining case i). In
this case the solution for $\phi_0$ is given exactly by
(\ref{sol}). The ansatz consistency condition and the remaining
equation of motion for $\A_i$ look as follows,
\begin{align}\label{remain1}
    &[\A_i,\phi_0]=0,\\ \label{remain2}
    &[\A_i,[\A_i,\A_j]]=0.
\end{align}

Let $\Xi_0$ denote the finite dimensional subspace of the Hilbert
space $\hh$, to which $\phi_0$ is projecting, $\dim \Xi_0=N $. And
denote the infinite dimensional orthogonal completion to $\Xi_0$
in $\hh$ as $\hh_0$. As an operator acting on $\hh$, $\phi_0$ is
the identity one when restricted to $\Xi_0$ and vanishes when
restricted to $\hh_0$. In a convenient basis it can be represented
in the following block matrix form,
\begin{equation}
   \phi_0=
   \begin{pmatrix}
     \I_0 & 0 \\
     0 & 0
   \end{pmatrix},
\end{equation}
where $\I_0$ stands for $N\times N$ unity matrix acting on
$\Xi_0$.

Eq. (\ref{remain1}) implies that gauge field $\A_i$ must have the
following block structure,\footnote{Note that from i) it follows
that the set of $\A_\mu$ cannot be irreducible, since $\phi_0$
commute with all $\A_\mu$ and is not a constant.}
\begin{equation}
   \A_i=
   \begin{pmatrix}
     B_i & 0 \\
     0 & C_i
   \end{pmatrix},
\end{equation}
where $B_i$ is $N\times N$ hermitian matrix acting on $\Xi_0$, and
$C_i$ is respectively Hermitian operator acting on (infinite
dimensional) Hilbert space $\hh_0$.

In terms of $B_i$ and $C_i$ the eq. (\ref{remain2}) is rewritten
as, two sets of independent equations,
\begin{align}\label{eq:b}
   & [B_i,[B_i,B_j]]=0 \\ \label{eq:c}
   & [C_i,[C_i,C_j]]=0
\end{align}

The general solution for the finite dimensional part (\ref{eq:b})
is given by a set of commuting matrices
\cite{Sochichiu:2000ud,Sochichiu:2000bg}, while for the eq.
(\ref{eq:c}), there are known various solution in different
dimensions \cite{Nekrasov:1998ss,Aoki:1999vr} (see also a recent
paper \cite{Belhaj:2000bs}). Let us only note that the one of the
simplest solutions is given by operators $C^{(0)}_i$ satisfying,
\begin{equation}
   [C^{(0)}_i,C^{(0)}_j]=\ii \theta_{ij},
\end{equation}
for some constant matrix $\theta_{ij}$.

In this case fields $C^{(0)}_i$ are generating a flat
noncommutative space corresponding to the ``equipotential''
surfaces of the solitonic field $\phi_0=\const$.

% ----------------------------------------------------------------
\section{Conclusions}
% ----------------------------------------------------------------
In this paper we have considered the noncommutative tachyonic
field interacting with noncommutative U(1) Yang--Mills model,
which is assumed to implement the dynamical noncommutativity.

We have shown that in this case instead of taking the limit of the
large noncommutativity parameter $\theta$, (in order to get rid of
kinetic term,) one can try find the gauge field background in
which the kinetic term vanishes naturally.

We reviewed a number of simple ans\"{a}tze which solve this
condition and have found that the only consistent one is the
simplest one in which the solitonic field is gauge covariant. The
existence of the more general ans\"{a}tze and solutions to the
equations of motion, including solitonic ones, also cannot be
ruled out. The last point can serve as a topic for future
investigations.

Another extension of this work can be seen in the introduction of
the supersymmetry. Let us note that the collective coordinate
$B_i$ arisen in the decomposition of gauge field $\A_i$ with
respect to projector $\phi_0$, satisfies the (bosonic part) of the
IKKT equations of motion for finite $N$. These may serve as an
indication to a deeper relation of the IKKT model with the
dynamics of unstable D-branes.

Since in the $N\to\infty$ limit of eq. (\ref{eq:b}) may possess
additional solutions beyond the commutative ones, it would be of
interest to analyse also this limit.

Finally, let us note that the presence of the noncommutative U(1)
gauge field makes possible the extension of the results of the
Ref. \cite{Sochichiu:2000bg}, concerning the equivalence of the
noncommutative models in different dimensions also to the
tachyonic system.
\subsection*{Note Added:}When this work was finished the same day appeared a
paper \cite{Gopakumar:2000rw}, containing analogous proposal to
introduce non-trivial gauge field interacting with tachyonic
field.
% ACKN -----------------------------------------------------------
\acknowledgments

The author is grateful for the hospitality to the Abdus Salam
International Center for Theoretical Physics, where this work has
been done. I would like to thank J.H.~Schwarz for pointing my
attention to the existing problem. I am grateful to Mounia
Hssaini for the help in the preparation this publication and also
thank her together with M.B.~Sedra for valuable discussions.

% BIBLIOGRAPHY ---------------------------------------------------
%\bibliographystyle{h-elsevier2}
%\bibliography{tachyon}

\end{document}